*A report of the BAA Variable Star Section*

**Unusual "stunted" outbursts in the nova-like variable HS 0229+8016**

Jeremy Shears

**Abstract**

We present the light curve of HS 0229+8016 between 2006 and 2019 which shows that the star varies between mag 13.4 and 15.1. There were episodes lasting up to hundreds of days during which distinctive 0.6 mag cyclic variations were apparent, each lasting ~9.5 days. Observationally, these are very similar to the "stunted" outbursts seen from time to time in several nova-like cataclysmic variables, such as UU Aqr. There are two intervals of up to 166 days when these stunted outbursts appeared to reduce or even stop. These coincide with the system being up to ~0.7 mag fainter than usual. Previous workers have suggested that HS 0229+8016 might be a member of the Z Cam family of dwarf novae, but we can find no evidence of the characteristic dwarf nova outbursts and standstills at an intermediate brightness. Surveillance of HS 0229+8016 should be continued to understand more about the long-term behaviour of this system.

**Introduction**

Cataclysmic variable stars (CVs) are binary stars in which a secondary donor star transfers gas to a companion white dwarf (1) (2). In the absence of a strong magnetic field the transfer occurs via an accretion disc. Some CVs are dwarf novae which undergo quasi-periodic outbursts of the accretion disc. Others, such as the nova-like CVs, have stable non-outbursting discs. (Most classical nova observed decades after the nova outburst appear similar to nova-like CVs). Most CVs with orbital periods below about 2.5h are reasonably well-behaved dwarf novae, but at longer periods there is a wide variety of poorly understood CVs which exhibit a range of behaviours.

HS 0229+8016 was identified as a CV during follow up observations of optically selected CV candidates from the Hamburg Quasar Survey (HQS) by Aungwerojwit *et al*. (3) (4). Its orbital period of 232.550 ± 0.049 min (0.16149 d) was determined from radial velocity and photometric variability studies. HS 0229+8016 exhibited clean quasi-sinusoidal radial velocity variations of its Hα emission line, but no or very little orbital photometric variability. They suggested that it is either a nova-like CV or a Z Cam-type dwarf nova which they just happened to observe always in outburst or standstill. The authors concluded their paper by saying "optical long-term monitoring will be necessary to distinguish between these two possibilities".

During discussions with the author in 2006, Boris Gänsicke suggested that he might like to carry out a long-term photometric monitoring of these stars; Gänsicke also set up a website listing the details of a number of CVs, including HS 0229+8016, which had been identified in the HQS (5). The results of the 13-year study are presented here.

The position and other details of HS 0229+8016 are given in Table 1.

**Observations**

To produce a long-term light curve of HS 0229+8016, we used data from the BAA VSS database and the AAVSO International Database and the observers are shown in Table 2. These data are mainly CCD measurements, either unfiltered or V-band using V-band comparison stars, although some visual estimates are included. The database includes some measurements with other filters, but these were not used in this study, the exception being a small number of TG measurements (tricolour digital imaging using data from the Green channel only). We also used V-band photometry from the All Sky Automated Survey for Supernovae (ASAS-SN), which is an automated sky survey to search for supernovae and other astronomical transients (6) (7) (8).

The AAVSO data also include several time-resolved photometric observations in addition to single point data (Table 3). These were used to investigate short-term variability of HS 0229+8016.

Encouraged by Boris Gänsicke, we also initiated an observational campaign at the end of 2018 November under the auspices of the BAA Variable Star Section (VSS). This was promoted via the BAA website, the BAAVSS-alert and CVnet-outburst email lists and the AAVSO Forum. The aim was to intensify the photometric coverage of the object's short-term behaviour, so observers were encouraged to obtain nightly snapshot measurements. Initially the campaign was intended to run until the end of 2019 February, but several observers continued to submit observations after that time.

**Results**

The thirteen-year light curve of HS 0229+8016 between 2006 April 22 and 2019 Jul 18 is presented in Figure 1. This shows considerable variation in brightness between mag 13.4 and 15.1, which we believe is inherent to the star. Several fading episodes are also apparent. The first of these took place in early 2012 and began around JD 2455968, when the star was mag ~14.5. It gradually faded over the next 90 days to mag 15.1. This was followed by a somewhat faster rise, over 40 days, to mag 14.3 on JD 2456058. Thus, the fade lasted ~90 days and was 0.6 to 0.8 mag in depth.

Another fading episode took place in late 2016/early 2017. It started around JD 2457694 when the star was mag ~14.1. The fade to a minimum mag ~14.8 took some 40 days and the recovery to mag ~14.0 took another 63 days. Thus, this fade lasted ~ 103 days and was 0.7- 0.8 magnitudes deep. The duration and depth of the two fades was therefore similar. Note that the declining part of the second even appeared to take place in two parts: a gradual stage, followed by a more rapid component.

Cursory analysis of the long-term light curve in Figure 1 also shows a third fading event during 2018 October, centred on JD 2456575. Expanding this section of the light curve reveals that this was a very brief episode lasting no more than 10 days during which the star faded from mag 14.4 to 15.1, or a depth of 0.7 magnitudes.

To clarify the variations in brightness between mag 13.4 and 15.1 mentioned above, we replotted the light curve from Figure 1 on a much expanded scale. The new light curve is shown in sections in Figure 2, where we no longer distinguish ASAS-SN data from the rest of the photometry. We decided to add lines joining each data point to guide the eye since without these lines it is troublesome to see whether there is a regular variation or merely scatter. What is revealed is that for much of the time there are continuous low amplitude cyclic variations. The episodes of continuous cycling behaviour lasted for up to 250 days (e.g. from 2456849 to 2457099) and in most cases probably continued for much longer, but gaps in the observational record mean it is not possible to say for how long.

Where possible, depending on the data cadence, the time of maximum of each cycle, its duration, amplitude, maximum brightness and the time between consecutive maxima were extracted. The data for the 85 cycles which were measured are listed in Table 4 and shown in Figure 1b , c, d. In some cases it was not possible to make reliable measurements either due to insufficient data cadence or because the cycles were not apparent; in other cases, maximum limits on the duration of the cycle were noted when too few data points defined the cycle. The mean duration was 9.5 days

(range 5 to 13 days) and the mean amplitude was 0.6 mag (range 0.2 to 0.9 mag). The mean interval between each maximum was 14.3 days, with a standard deviation of 4.2 days; the shortest time between brightening events was 4.1 days and the longest 27.6 days. We note that in the later stages of the observational window, after approximately JD 2458500, the cycle duration, amplitude and time between successive maxima were all decreasing (Figure 1b , c, d).

There are two intervals when these low amplitude cycles appeared to reduce or even stop. The first was between JD 2455916 and 2456068 (2011 Dec 20 to 2012 May 20), i.e. about 92 days. The second was between JD 2457671 and 2457837 (2016 Oct 9 to 2017 Mar 24), 166 days. Each episode might have been longer, but the lack of data outside these interval means they cannot be constrained further. This coincides with the first two fading episodes mentioned above.

Time resolved photometry was carried out by several observers in 2014 October and during the declining branch of the 2016 fading event. An observing log is shown in Table 3 and light curves of some of the longer runs are shown in Figure 3. These plots clearly show short-period variability with an amplitude of ~ 0.05 to 0.1 mag, which appear to be quasi-periodic variations on time scales of ~ 20 – 40 min and some at 2 to 4 h. However, no consistently repeating variation is detected; neither is there evidence of eclipses despite the star being observed over multiple orbital periods. These variations are typical of the flickering seen in many CVs. We analysed the combined datasets for the time-resolved photometry runs, as well as the daily runs, using the Lomb–Scargle and ANOVA algorithms in the *Peranso* V2.50 software (9), but could not find a significant stable period in the power spectra, including in the region of the proposed orbital period (data not shown). The absence of a signal associated with the orbital period is consistent with the small data set that Aungwerojwit *et al.* (4) presented.

Discussion

The most remarkable aspect of the long-term light curve of HS 0229+8016 is the ~9.5-day, 0.6 mag cyclic variations during much of the light curve. The cyclic events appear similar to dwarf nova outbursts in duration but are of much smaller amplitude. These are very similar to the 0.4 - 1 mag variations seen periodically in a few nova-like CVs, such as UU Aqr, Q Cyg, CP Lac, W Ser and RW Sex, which Honeycutt calls "stunted" outbursts (10) (11) (12). For example, UU Aqr is a nova-like system which shows 0.6 mag stunted outbursts which have durations of about 10 days, which is remarkably similar to HS 0229+8016. This similarity also extends to their orbital period, which is 3.89 and 3.93 h respectively. Moreover, UU Aqr exhibits stunted outbursts for extended intervals, but at other times they are missing in much the same way as we found for long intervals when the stunted outbursts in HS 0229+8016 reduced or even stopped.

Some of the nova-like systems which exhibit stunted outbursts also display dips in their light curve, which are shaped much like inverted stunted outbursts (12). These dips are less numerous and more diverse than stunted outbursts. Both isolated dips and adjacent pairs of outbursts and dips are found – see for example Figure 4 which shows outburst/dip pairs in the nova-like CVs, RW Sex and Q Cyg. Intriguingly, Figure 5 shows examples of similar dips in HS 0229+8016.

The mechanism for stunted outbursts remains uncertain. There is evidence that they are due to mass transfer events (10) (13) and also evidence that they are due to accretion disc instabilities occurring under unusual circumstances (11), perhaps indicating that only part of the accretion disc goes into outburst. Therefore, it might be important to note that the times when the stunted outbursts of HS 0229+8016 decreased or even ceased corresponded to times when the system was undergoing one of its fading episodes. Since much of the light output of a CV is contributed by the accretion disc, it is possible that the quantity of material present in the disc during these fades is somewhat reduced, thereby rendering the disc less likely to become unstable and hence less likely to exhibit stunted outbursts. On the other hand, according to classical disc in stability theories of dwarf novae, a drop in the brightness should reflect a drop in the accretion rate, which would increase the chance that parts of the disc undergo outbursts. Further studies on HS 0229+8016 may help to understand these stunted outbursts better.

Though possibly not physically related to the stunted outbursts reported here, some VY Scl stars show what appear to be mini-outbursts during their low states. This includes MV Lyr which shows quasiperiodic (~2h) brightening pulses during its low state which are thought to be controlled by the magnetic field of the white dwarf (14).

Another class of CV which undergoes almost continuous cycles of outbursts for extended periods of time is the Z Cam family of dwarf novae. Z Cam stars alternate between episodes of continuous dwarf nova-type outbursts and standstills during which the light curve is similar to stable nova-like systems. Furthermore, some Z Cam systems display outbursts/dip pairs similar to those reported for UU Aqr and other nova-like systems (12) and HS 0229+8016 as discussed above. Examples of such Z Cam systems include Z Cam itself and AH Her. However, there are important differences between stunted outbursts and Z Cam outbursts, the most important of which is that the Z Cam stars typically have 2.5 mag or greater outbursts (outside standstills, of course). Also, the defining characteristic of Z Cams is the standstills: times when the star gets stuck at a mid-point between maximum and minimum. The episodes during which HS 0229+8016 ceases to have stunted outburst are qualitatively different from Z Cam-like standstills, not least because the system is fainter than normal. On this basis there is no evidence from the present study to support HS 0229+8016 being a member of the Z Cam family. Having said that, the classification of Z Cam stars is not astrophysically based, rather it is merely on the basis of a star's light curve. The light curve should show a repeated cycling between proper dwarf nova outbursts, standstills with no significant variability and a mean magnitude in between the quiescent and peak magnitude during the outbursting stage. When restricting Z Cam to this type of behaviour, these systems can be explained by having an average accretion rate that is close to the critical rate to have unstable discs.

Finally, it is worth noting that the orbital period range of the CVs which exhibit the various phenomena discussed in this paper (stunted outbursts, VY Scl-like low states, SW Sex phenomena, Z Cam behaviour) at least superficially appear to cluster in the 3-4h range. It may therefore be that these behaviours are interrelated.

This work shows once again the importance of amateur astronomers maintaining a long-term record of a star's brightness and also of contributing to a coordinated campaign. It is clear from the light curves presented here that the campaign initiated in 2018 November resulted in the most intensive coverage of the system, greatly enhancing the quality of the light curve. It is hoped that observers will continue to monitor this star, in particular to see if the decrease in the stunted outburst cycle duration, amplitude and time between successive maxima noted above will continue.

**Conclusions**

The light curve of HS 0229+8016 between 2006 and 2019 shows that the star varies between magnitude 13.4 and 15.1. The striking feature of the light curve were episodes during which there were almost continuous 0.6 mag cyclic variations, each cycle had a duration of ~9.5 days, lasting for hundreds of days. These cycles appear similar to dwarf nova outbursts in duration, but they have a much smaller amplitude. Observationally, they are very similar to the "stunted" outbursts seen periodically in many nova-like CVs, such as UU Aqr which has a very similar orbital period to HS0229+8016. There are two intervals of up to 166 days when these stunted outbursts appeared to reduce or even stop. These coincide with the system being up to ~0.7 mag fainter than usual.

Previous workers have suggested that HS 0229+8016 might be a member of the Z Cam family of dwarf novae, but we can find no evidence of the characteristic dwarf nova outbursts, nor of standstills which such a designation would require.

Surveillance of HS 0229+8016 should be continued to understand more about the long-term behaviour of this CV.

**Acknowledgments**

The author acknowledges with thanks the use of photometry from the databases of the BAA VSS, the AAVSO and the All Sky Automated Survey for Supernovae. The research made use of the NASA/Smithsonian Astrophysics Data System and SIMBAD, operated through the Centre de Données Astronomiques (Strasbourg, France).

The author is grateful to Professor Boris Gänsicke, University of Warwick, not only for encouraging him to take on this project, but also for sharing his enthusiasm and knowledge over many years. I would especially like to thank those observers who contributed to the intensive campaign launched in 2018 November.

I thank both referees for their helpful comments which have improved the paper.

| Constellation | Cepheus |
|---|---|
| RA, Dec (J2000.0) | 02 35 58.23, +80 29 44.2 |
| Other names | 2MASS J02355820+8029441<br>GSC 04503-00345<br>UCAC4 853-003056 |
| Orbital period | 0.16149 d<br>3.8758 h |

**Table 1: Information on HS 0229+8016**

| Observer | Country | Method |
|---|---|---|
| Armiński, Andrzej | PL | CCD + V |
| Boardman, James * | US | CCD + V or C |
| Boyd, David * | GB | CCD + V |
| Cook, Lewis (Lew) | US | CCD + C |
| de Miguel, Enrique | ES | CCD + V |
| Dufoer, Sjoerd * | BE | CCD + V |
| Fleming, George * | GB | CCD + V |
| Johnston, Steve | GB | CCD + V |
| Joslin, Mel * | GB | Vis |
| Kautter, William * | US | CCD + V |
| Menzies, Ken * | US | CCD + V |
| Miller, Ian * | GB | CCD + V |
| Mobberley, Martin * | GB | CCD + V |
| Morales Aimar * | ES | CCD + V |
| Morelle, Etienne | FR | CCD + C |
| Padovan, Stefano | ES | CCD + V |
| Popov, Velimir | BG | CCD + V |
| Poyner, Gary * | GB | CCD + V, Vis |
| Reichenbacher, Frank | US | Visual |
| Roe, James | US | CCD + V |
| Sabo, Richard * | US | CCD + V |
| Schwendeman, Erik * | US | CCD + V |
| Shears, Jeremy * | GB | CCD + C |
| Smith, Dave * | GB | CCD + V |
| Simonsen, Michael | US | Vis |
| Staels, Bart | BE | CCD + V |
| Stein, William | US | CCD + V |
| Stone, Geoffrey | US | CCD + C |
| Storey, David * | GB | CCD + V |
| Swan, David * | GB | CCD + V or TG |
| Szaj, Robert | PL | Vis |
| Taylor, Daniel | CA | Vis |
| Tomlin, Ray * | US | CCD + V |
| Tordai, Tamas * | HU | CCD + V |
| Vanmunster, Tonny * | BE | CCD + V |
| Zecchin, Franck | FR | CCD + V |

**Table 2: HS 0229+8016 observers**

CCD + V = CCD with V filter; CCD + C = unfiltered CCD; TG = tricolour imaging using Green channel only

* indicates observers who also contributed to the VSS campaign from 2018 November.

| Observer | Date | JD start | JD end | Duration (h) | Mean magnitude |
|---|---|---|---|---|---|
| Stein | 2014 Oct 29 | 2456959.575 | 2456960.021 | 10.7 | 14.25 |
| Stein | 2014 Oct 30 | 2456960.626 | 2456961.022 | 9.5 | 14.19 |
| Cook | 2016 Nov 24 | 2457716.642 | 2457716.809 | 4.0 | 14.52 |
| Stone | 2016 Nov 25 | 2457717.586 | 2457718.031 | 10.7 | 14.42 |
| Boardman | 2016 Nov 26 | 2457719.492 | 2457719.688 | 4.7 | 14.64 |
| Cook | 2016 Nov 29 | 2457721.679 | 2457721.892 | 5.1 | 14.60 |
| Morelle | 2016 Dec 1 | 2457724.220 | 2457724.647 | 10.2 | 14.68 |
| Cook | 2016 Dec 1 | 2457724.630 | 2457724.859 | 5.5 | 14.62 |
| Morelle | 2016 Dec 2 | 2457725.242 | 2457725.599 | 8.6 | 14.70 |
| Cook | 2016 Dec 4 | 2457726.650 | 2457726.817 | 4.0 | 14.57 |
| Morelle | 2016 Dec 10 | 2457733.223 | 2457733.624 | 9.6 | 14.75 |
| Morelle | 2016 Dec 22 | 2457745.291 | 2457745.355 | 1.5 | 14.74 |
| Morelle | 2016 Dec 23 | 2457746.231 | 2457746.331 | 2.4 | 14.79 |

**Table 3: Time resolved photometry log**

| JD | Duration (days) | Amplitude (mag) | Max brightness (mag) | Time since previous max (d) |
|---|---|---|---|---|
| 2456182.8 | <12 | 0.7 | 13.9 | |
| 2456200.0 | <17 | 0.7 | 13.8 | 17.2 |
| 2456216.9 | 10 | 0.7 | 13.9 | 16.9 |
| 2456234.4 | <17 | 0.6 | 13.9 | 17.5 |
| 2456252.8 | <13 | 0.7 | 13.9 | 18.4 |
| 2456267.8 | 11 | 0.7 | 13.8 | 15.0 |
| 2456286.4 | <14 | 0.8 | 13.8 | 18.6 |
| 2456304.3 | <12 | 0.5 | 14.0 | 17.9 |
| 2456316.6 | <17 | 0.5 | 13.9 | 12.3 |
| 2456334.6 | 12 | 0.7 | 13.8 | 18.0 |
| 2456354.6 | <15 | 0.8 | 13.9 | 20.0 |
| 2456366.4 | <13 | 0.7 | 13.9 | 11.8 |
| 2456386.4 | <13 | 0.7 | 13.9 | 20.0 |
| 2456400.7 | <12 | 0.8 | 13.8 | 14.3 |
| …….. | …….. | …….. | ……. | …….. |
| 2456515.5 | <13 | 0.7 | 13.7 | |
| 2456532.6 | 10 | 0.7 | 13.8 | 17.1 |
| 2456550.4 | 10 | 0.7 | 13.8 | 17.8 |
| …….. | …….. | …….. | ……. | …….. |
| 2456599.9 | <16 | 0.9 | 13.6 | |
| 2456619.4 | <15 | 0.7 | 13.8 | 19.5 |
| …….. | …….. | …….. | ……. | …….. |
| 2456649.4 | <12 | 0.7 | 13.6 | |
| 2456667.3 | <13 | 0.9 | 13.6 | 17.9 |
| 2456682.7 | <13 | 0.9 | 13.6 | 15.4 |
| 2456698.4 | <14 | 0.7 | 13.7 | 15.7 |
| 2456714.4 | 11 | 0.7 | 13.8 | 16.0 |
| 2456730.3 | 11 | 0.7 | 13.8 | 15.9 |
| 2456746.3 | 11 | 0.6 | 13.8 | 16.0 |
| 2456760.3 | 9 | 0.6 | 13.8 | 14.0 |
| 2456779.3 | 9 | 0.5 | 13.8 | 19.0 |
| …….. | …….. | …….. | …….. | …….. |
| 2456849.1 | <17 | 0.5 | 14.0 | |
| 2456866.9 | <14 | 0.7 | 13.8 | 17.8 |
| 2456879.7 | 12 | 0.5 | 13.9 | 12.8 |
| 2456895.1 | 11 | 0.7 | 13.7 | 15.4 |
| 2456908.1 | 11 | 0.9 | 13.6 | 13.0 |
| 2456922.4 | 9 | 0.7 | 13.8 | 14.3 |
| 2456935.7 | 11 | 0.6 | 13.8 | 13.3 |
| 2456951.6 | <14 | 0.5 | 13.9 | 15.9 |
| 2456965.4 | 11 | 0.4 | 13.8 | 13.8 |
| 2456978.9 | 12 | 0.9 | 13.7 | 13.5 |
| 2456993.9 | 11 | 0.5 | 13.8 | 15.0 |
| 2457005.9 | <13 | 0.9 | 13.7 | 12.0 |
| 2457017.9 | 10 | 0.5 | 13.7 | 12.0 |
| 2457030.7 | 11 | 0.5 | 13.8 | 12.8 |
| 2457044.8 | 12 | 0.6 | 13.7 | 14.1 |
| 2457057.4 | 9 | 0.4 | 13.8 | 12.6 |
| 2457072.8 | 10 | 0.4 | 13.7 | 15.4 |
| 2457085.4 | 11 | 0.4 | 13.9 | 12.6 |
| 2457099.3 | ND | 0.5 | 13.8 | 13.9 |
| …….. | …….. | …….. | …….. | …….. |
| 2457985.5 | 11 | 0.5 | 13.8 | |
| 2458013.1 | <19 | 0.7 | 13.7 | 27.6 |
| 2458029.0 | <15 | 0.8 | 13.5 | 15.9 |
| 2458046.0 | <19 | 0.5 | 13.6 | 17.0 |
| 2458062.9 | 13 | 0.6 | 13.6 | 16.9 |
| 2458079.0 | 12 | 0.6 | 13.6 | 16.1 |
| 2458097.0 | 10 | 0.5 | 13.5 | 18.0 |
| 2458117.8 | 11 | 0.4 | 13.6 | 20.8 |
| 2458138.3 | <14 | 0.7 | 13.6 | 20.5 |
| 2458152.9 | <12 | 0.7 | 13.6 | 14.6 |
| 2458167.7 | <12 | 0.8 | 13.4 | 14.8 |
| 2458191.3 | 11 | 0.5 | 13.7 | 23.6 |
| 2458200.6 | ND | 0.6 | 13.6 | 9.3 |
| …….. | …….. | …….. | …….. | …….. |
| 2458408.0 | 11 | 0.6 | 13.6 | |
| 2458420.3 | 11 | 0.7 | 13.5 | 12.3 |
| 2458435.9 | <13 | 0.6 | 13.6 | 15.6 |
| 2458442.9 | 9 | 0.4 | 13.9 | 7.0 |
| 2458453.4 | 7 | 0.4 | 13.9 | 10.5 |
| 2458462.9 | 8 | 0.4 | 13.9 | 9.5 |
| 2458469.8 | 8 | 0.4 | 13.9 | 6.9 |
| 2458475.9 | 7 | 0.4 | 13.9 | 6.1 |
| 2458490.5 | 8 | 0.4 | 13.9 | 14.6 |
| 2458498.5 | 8 | 0.4 | 13.9 | 8.0 |
| 2458507.4 | 7 | 0.4 | 13.9 | 8.9 |
| 2458516.0 | 8 | 0.4 | 13.9 | 8.6 |
| 2458523.6 | 7 | 0.6 | 13.9 | 7.6 |
| 2458532.7 | 7 | 0.4 | 13.9 | 9.1 |
| 2458547.6 | 7 | 0.6 | 13.9 | 14.9 |
| 2458558.3 | 9 | 0.7 | 13.8 | 10.7 |
| 2458567.3 | 5 | 0.7 | 14.0 | 9.0 |
| 2458571.4 | 8 | 0.5 | 13.9 | 4.1 |

| …….. | …….. | …….. | …….. | …….. |
|---|---|---|---|---|
| 2458608.4 | 6 | 0.4 | 13.8 | |
| 2458613.5 | ND | 0.2 | 14.1 | 5.1 |
| 2458624.4 | 6 | 0.4 | 13.9 | 10.9 |
| 2458637.6 | 7 | 0.5 | 13.9 | 13.2 |
| 2458649.6 | 8 | 0.5 | 13.9 | 12.0 |
| 2458661.6 | 8 | 0.6 | 13.9 | 12.0 |
| 2458677.6 | 10 | 0.5 | 14.0 | 16.0 |

**Table 4: Analysis of each cyclic variation**

ND: not determined

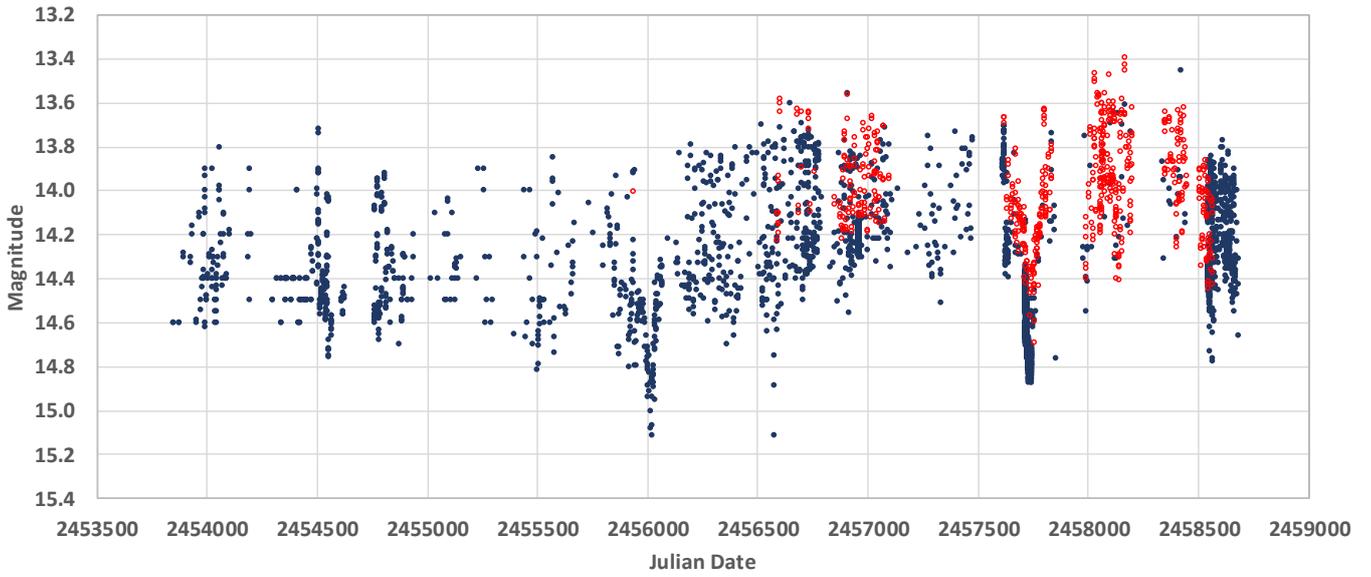
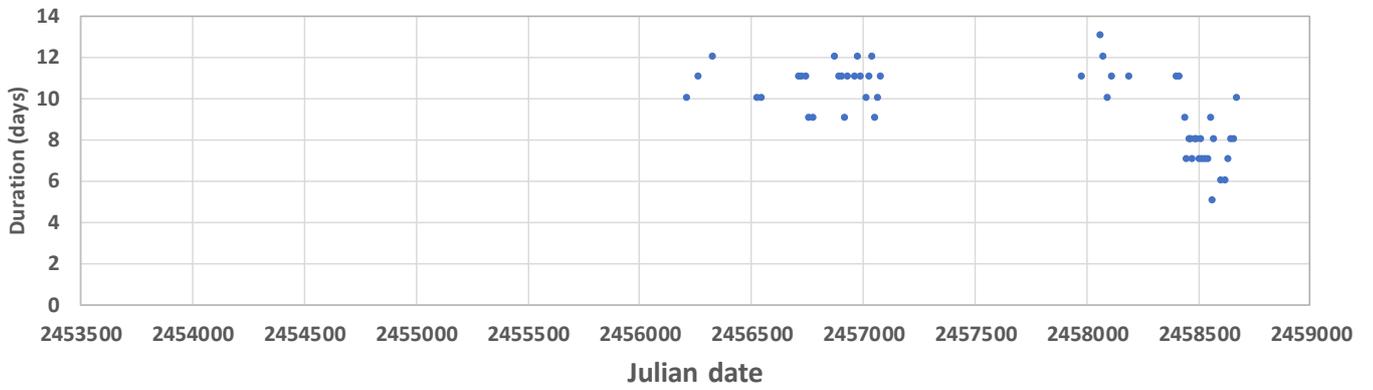
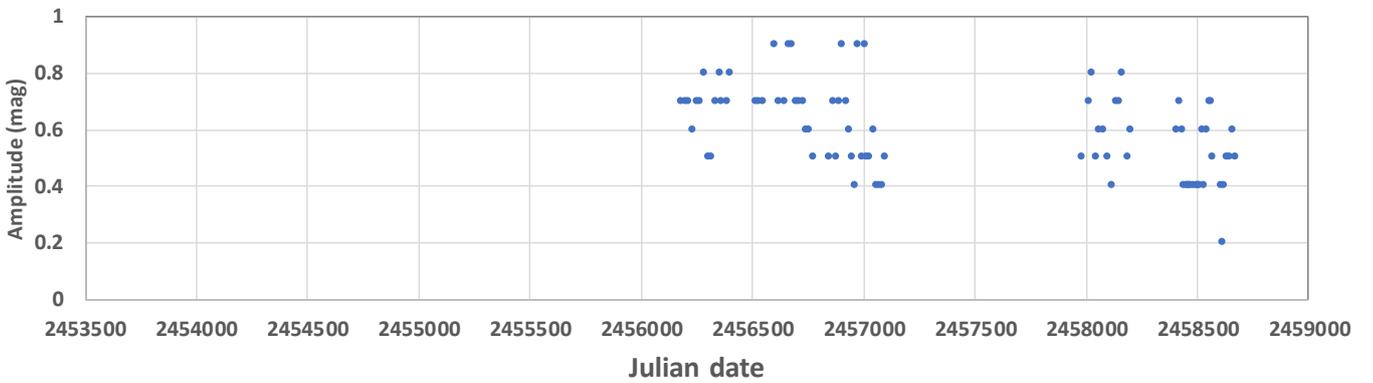
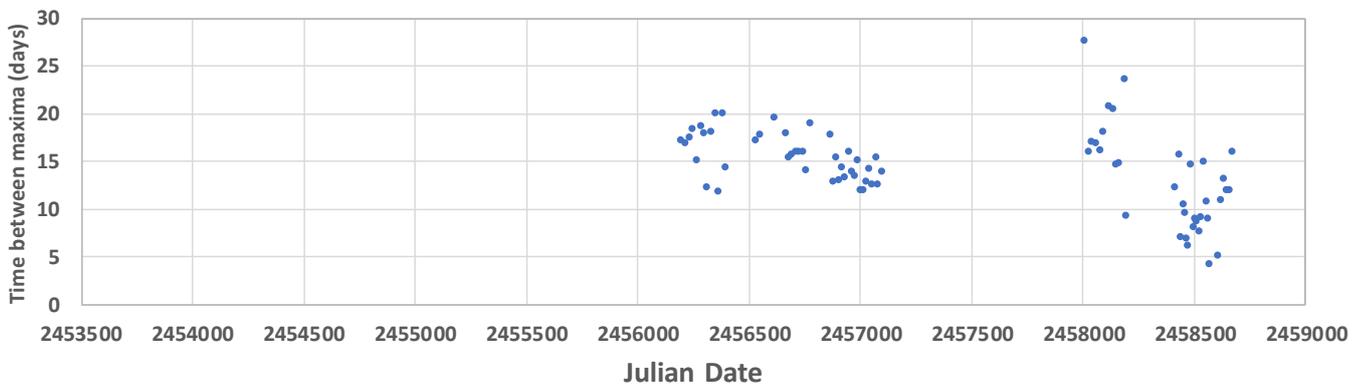

**Plots on previous page:**

**Figure 1 (a): Light curve of HS 0229+8016 between 2006 April 22 and 2019 Jul 18**

Blue data points: data from the BAA VSS database and the AAVSO International Database. Red data points: V-band photometry from ASAS-SN

**Fig 1 (b) Duration of each cyclic variation**

**Fig 1 (c) Amplitude of each cyclic variation**

**Fig 1 (d) Time between maxima of each successive cyclic variation**

Figure 2 **(on next two pages)**: The long-term light curve of HS 0229+8016

Each plot covers 300 days and shows the same data as in Figure 1. Lines have been added to guide the eye, making the cyclic variations discussed in the paper more obvious.

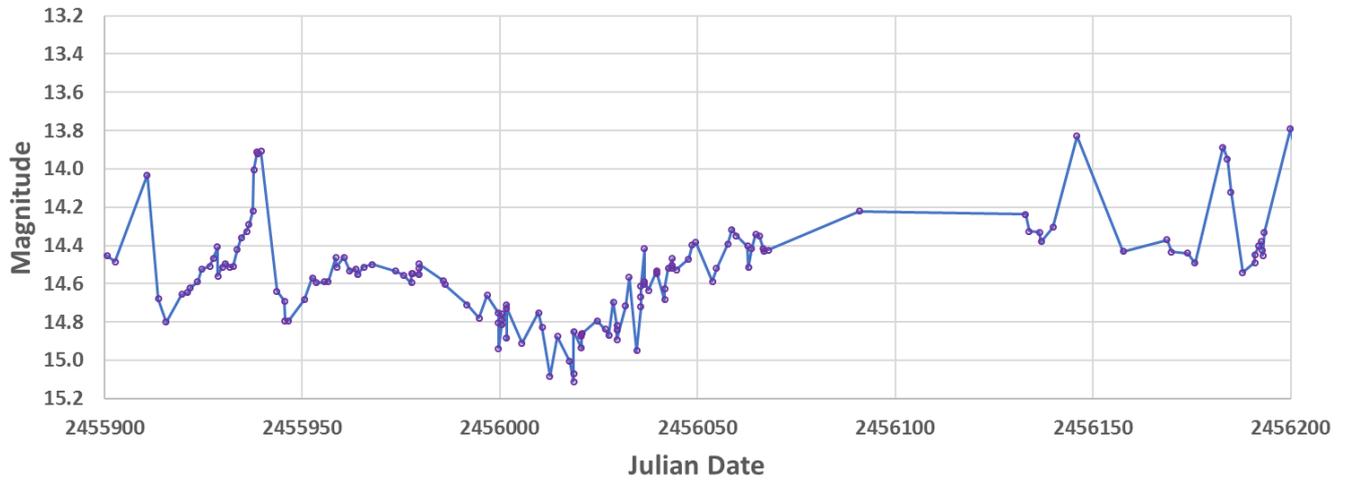
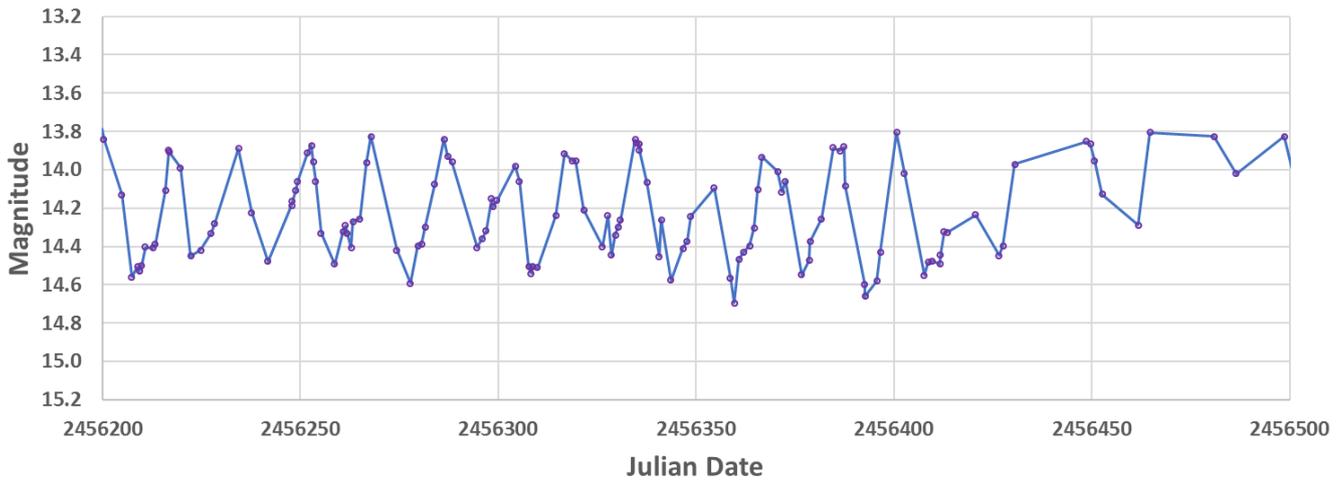
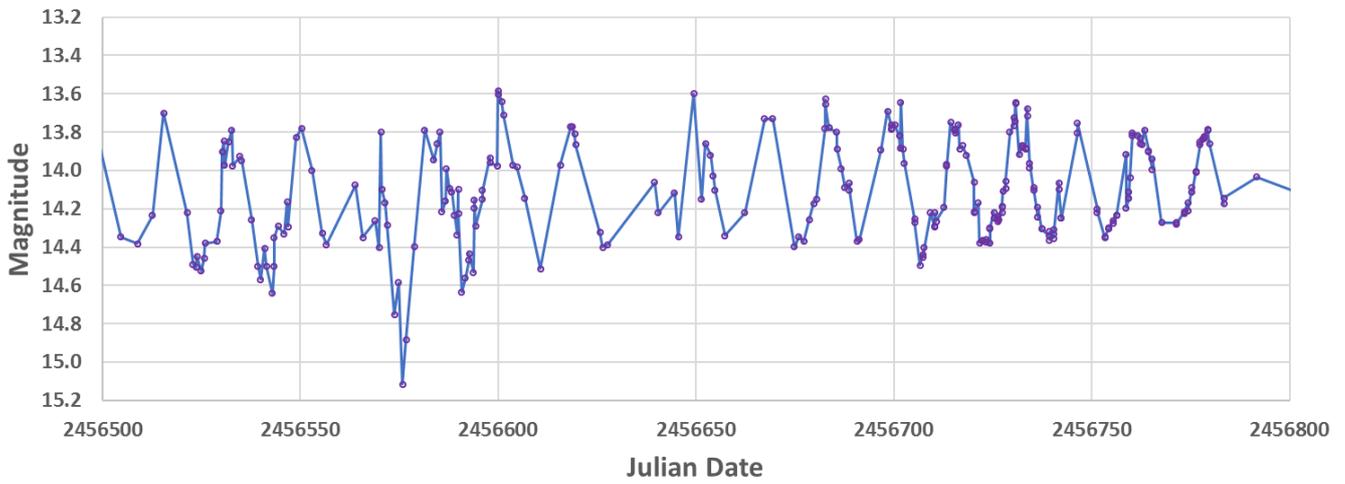
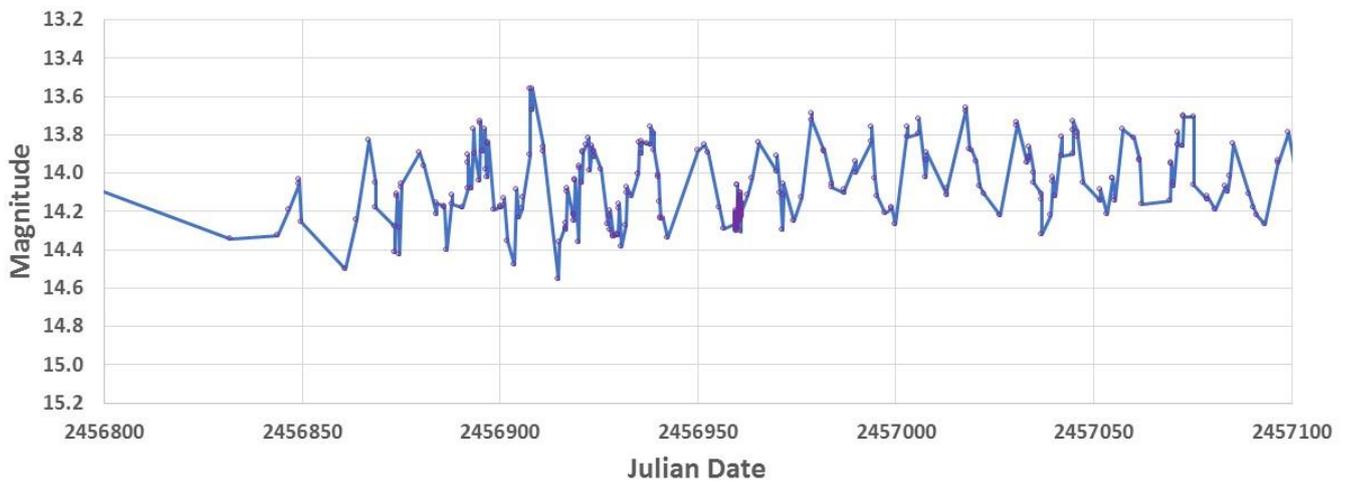

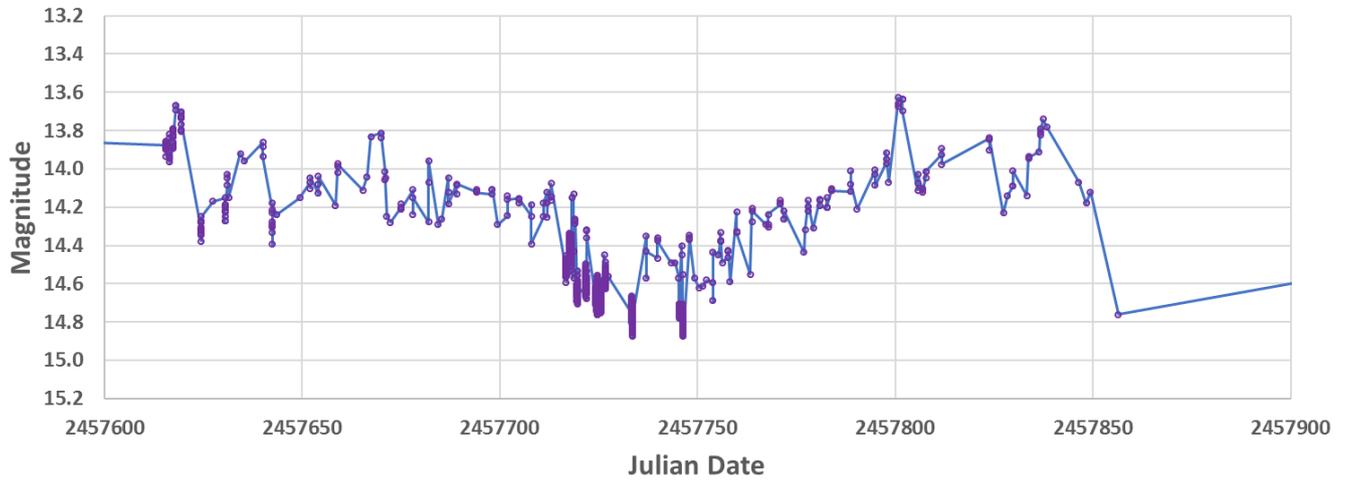
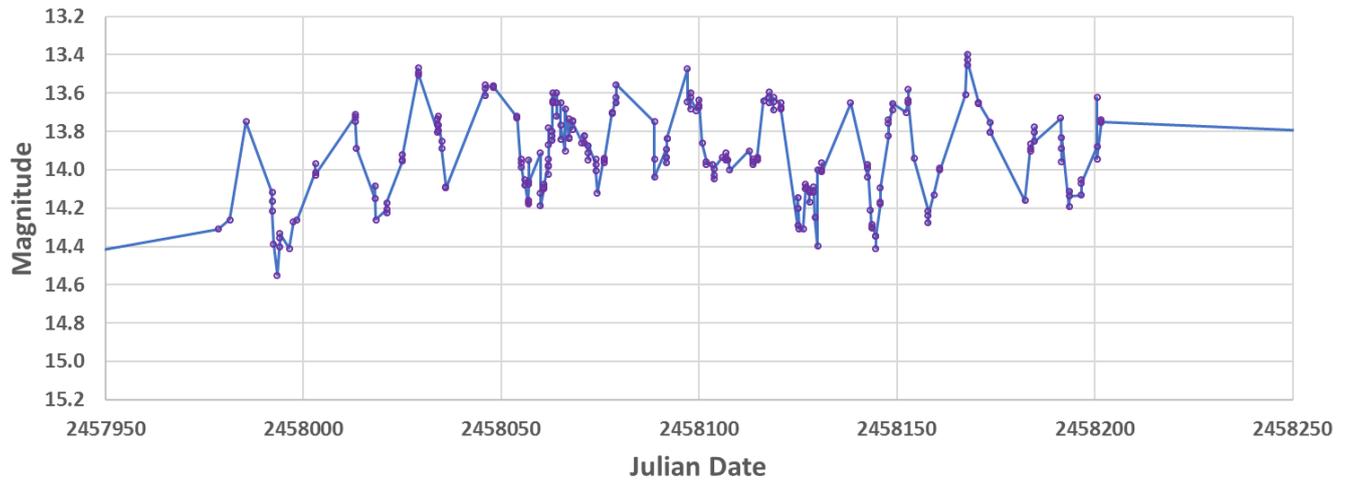
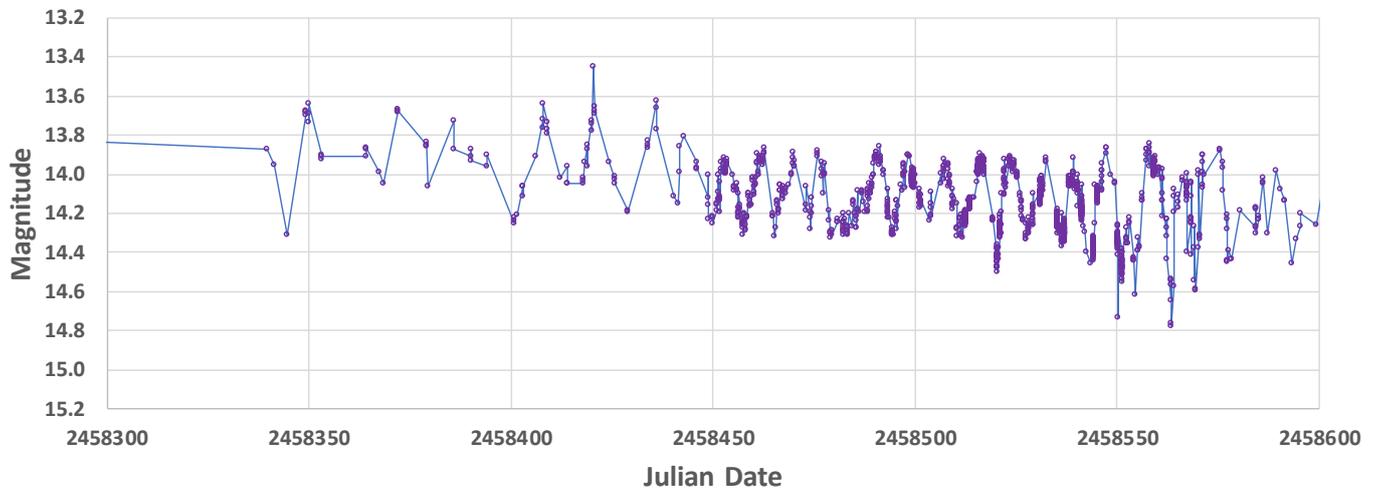
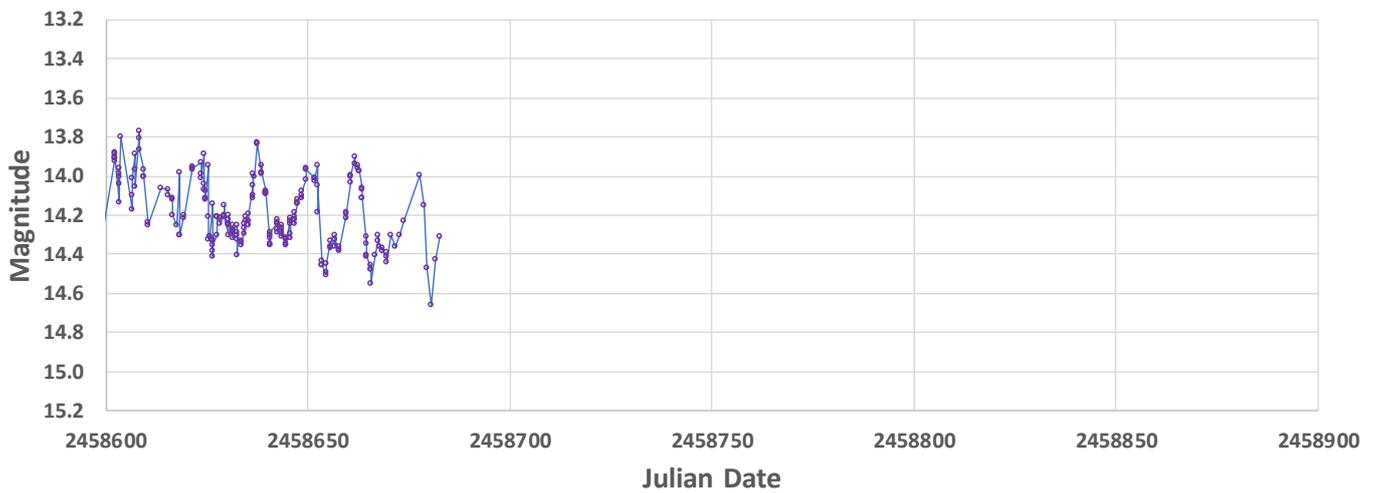

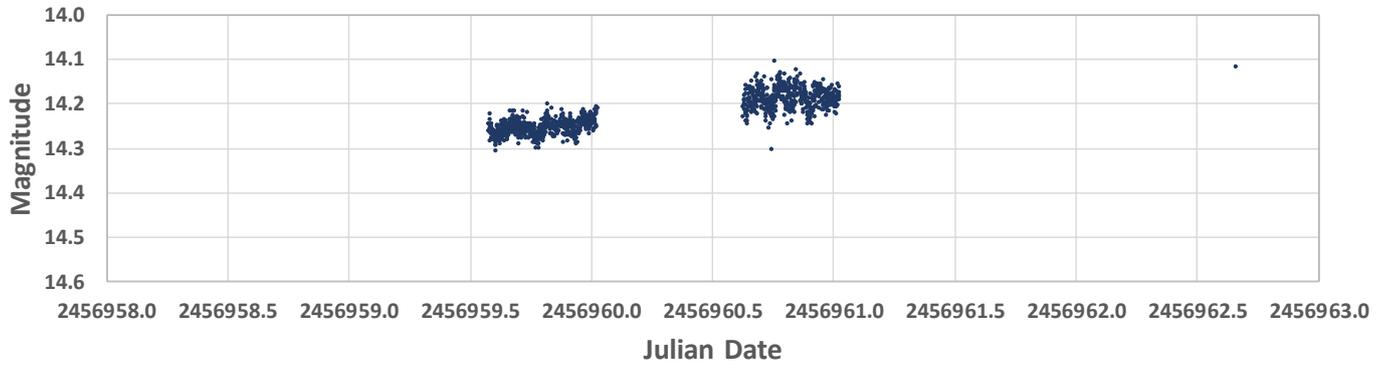
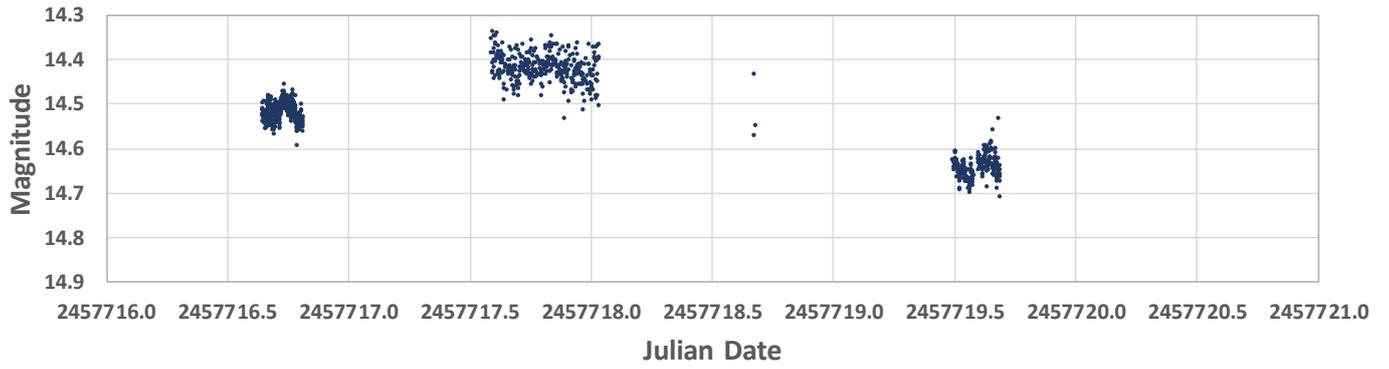
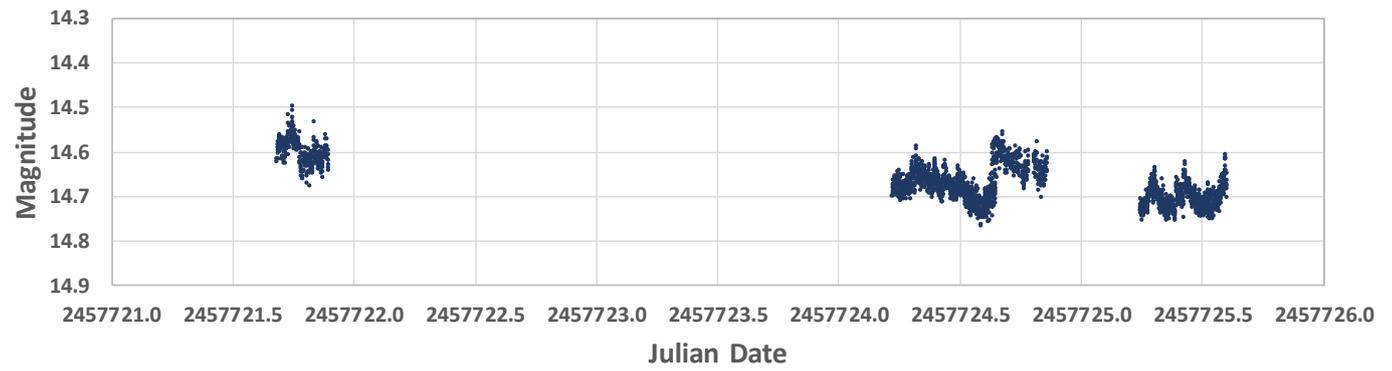
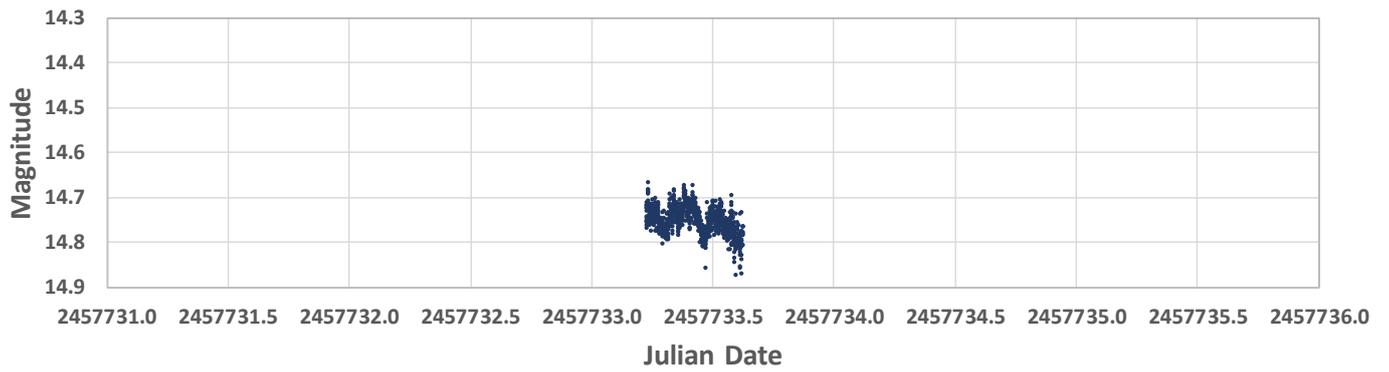

Figure 3: Examples of time-resolved photometry

Each plot shows an interval of 5 days (a) 2014 October, (b) - (d) 2016 November and December

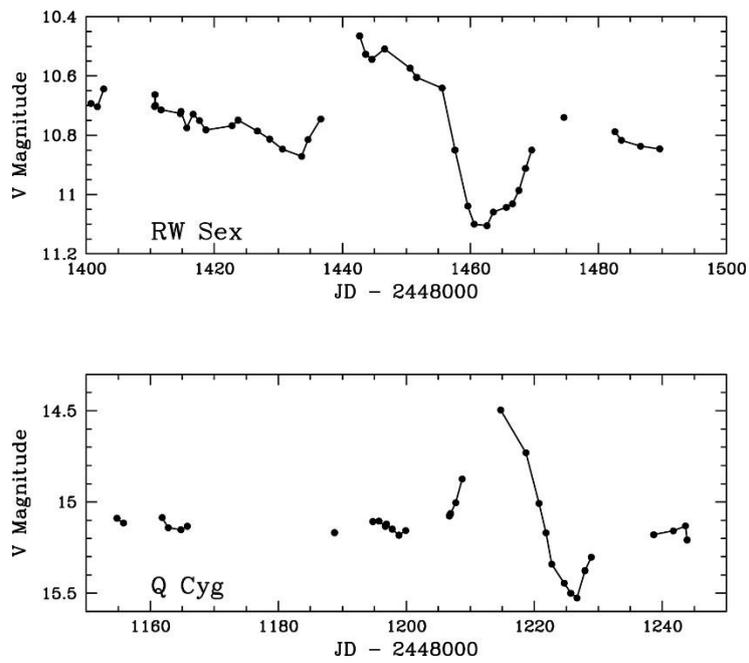

Figure 4: Stunted outburst/dip pairs in the nova-like CVs, RW Sex and Q Cyg

From reference (11)

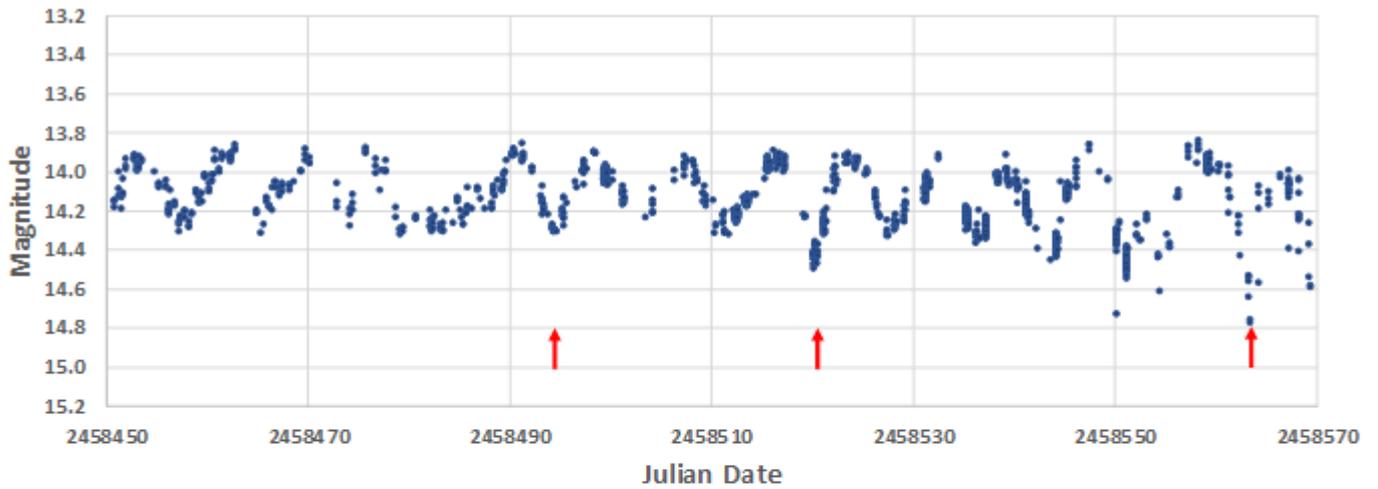

Figure 5: Examples of dips in HS 0229+8016